\documentclass[apl,a4paper,floatfix,twocolumn]{revtex4}
\usepackage[dvips]{graphicx}
\usepackage{amsfonts}
\usepackage{amssymb}
\usepackage{color}
\usepackage{setspace}

\begin{document}
\title{Andreev spectroscopy of CrO$_{2}$ thin films on TiO$_{2}$ and Al$_{2}$O$_{3}$}
\date{Published online 7th October 2013}
\author{K.A. Yates}
\affiliation{The Blackett Laboratory, Physics Department, Imperial}
\email[]{k.yates@imperial.ac.uk}
\author{M.S. Anwar}
\author{J. Aarts}
\affiliation{Physics Department, University of Leiden, The Netherlands}
\author{O. Conde}
\affiliation{Department of Physics and ICEMS, Faculty of Sciences of the University of Lisbon, Campo Grande, Ed. C8 1749-016 Lisboa, Portugal}
\author{M. Eschrig}
\affiliation{SEPnet and Hubbard Theory Consortium, Department of Physics, Royal Holloway, University of London, Egham, Surrey, UK, TW20 0EX}
\author{T. L\"{o}fwander}
\affiliation{ Department of Microtechnology and Nanoscience - MC2, Chalmers University of Technology, SE-412 96 G\"{o}teborg, Sweden}
\author{L.F. Cohen}
\affiliation{The Blackett Laboratory, Physics Department, Imperial
College, London, SW7 2AZ, UK}


\begin{abstract}
Here we analyse the spectroscopic information gathered at a number of single CrO$_{2}$ / Pb interfaces. We examine thin films requiring additional interfacial layers to generate long range spin triplet proximity effect superconductivity (CrO$_{2}$/TiO$_{2}$) or not (CrO$_{2}$/Al$_{2}$O$_{3}$). We analyse the data using two theoretical models and explore the use of a parameter free method to determine the agreement between the models and experimental observations, showing the necessary temperature range that would be required to make a definitive statement. The use of the excess current as a further tool to distinguish between models is also examined.  Analysis of the spectra demonstrates that the temperature dependence of the normalised zero bias conductance is independent of the substrate onto which the films are grown. This result has important implications for the engineering of interfaces required for the long range spin triplet proximity effect.
\end{abstract}
\maketitle

\section{Introduction}
In 2006 Keizer et al., reported the observation of a supercurrent transported over nearly a micron  through fully spin polarised \cite{Soulen, Yates, Lofwander} CrO$_{2}$ in an SFS Josephson junction \cite{Keizer}.   The conventional superconducting proximity effect is expected to be very short in CrO$_{2}$ \cite{EschrigPT} and certainly far shorter than the distances observed by Keizer et al. \cite{Keizer}.  The result though was explicable within the developing theory of the long range spin triplet proximity effect (LRSTPE) \cite{EschrigPT, EschrigNP, Bergeret}.  In this theory, two components are required to transform spin singlet Cooper pairs from the superconductor into spin parallel, triplet pairs that can exist in the ferromagnet: Spin mixing is required to mix the singlet pair ($|\uparrow \downarrow \rangle -|\downarrow \uparrow \rangle$) into a triplet pair of the form ($|\uparrow \downarrow \rangle +|\downarrow \uparrow \rangle$) while a further spin flip (or spin transformation) process is needed to change that spin opposite triplet pair component into the long range spin parallel ($|\uparrow \uparrow \rangle$ and $|\downarrow \downarrow \rangle$) one.  The theory predicted that magnetic inhomogeneity at the interface between a ferromagnet and a superconductor provided the key to both the spin mixing, defined by a spin mixing angle $\theta$ and spin transformation processes required for generation of the LRSTPE \cite{EschrigPT, EschrigNP, Bergeret}.

Subsequent experiments have confirmed that the inhomogeneity requirement for LRSTPE generation can be satisfied by engineered multilayer contacts \cite{Khaire} or intrinsic magnetic inhomogeneity such as the spiral ordering found in the rare earth metal holmium \cite{Robinson, Sosnin}.  Recent results on CrO$_{2}$ show that the LRSTPE can be observed in CrO$_{2}$ grown on Al$_{2}$O$_{3}$ substrates using simple superconducting contacts \cite{AnwarPRB} but only through CrO$_{2}$ films grown onto TiO$_{2}$ substrates when engineered contacts incorporating a thin Ni layer are employed \cite{AnwarAPL}.  Here we examine the spectroscopic information from single S/F, Pb/CrO$_{2}$ contacts using CrO$_{2}$ thin films grown on Al$_{2}$O$_{3}$ or TiO$_{2}$ substrates \cite{AnwarPRB, AnwarAPL}. We do this in the context of the Mazin modified Blonder-Tinkham-Klapwijk (BTK) \cite{BTK, Mazin} and the more recent LRSTPE spin mixing model (SMM) \cite{Lofwander, Grein} and find that the latter suggests that CrO$_{2}$ has strong intrinsic spin mixing properties independent of the substrate on which it is grown.
 
The spectroscopic method we use is point contact spectroscopy (PCS).  The technique has been used extensively to investigate the transport spin polarization of candidate materials for spintronics \cite{Soulen, Bugoslavsky} using the modified BTK model.   In this model, the spectra can be fitted using four parameters: the superconducting gap, $\Delta$, a measure of the interface scattering, $Z$ (a spin independent delta function parameter), the spin polarisation of the transport carriers, $P$ and either a spreading resistance, $r_{s}$, that captures the series resistance of the film as is used here \cite{Lofwander, Woods}, or a spectral broadening parameter that incorporates thermal and non-thermal smearing, $\omega$ \cite{Bugoslavsky}.   Within the modified BTK model, the spins are treated equally as they cross the interface and no consideration is given for effects such as spin dependent scattering \cite{Xia}.  It is clear though that for situations where there are strong spin mixing effects, the two spins will conduct differently across the interface.  The SMM model has been proposed to account for these differences.  In such cases, the conductance spectra observed by PCS will differ considerably from the predictions of the modified BTK model \cite{Lofwander, Piano}.  In reference \cite{Lofwander}, two independent measures were proposed to distinguish between the two models: Using the variation of the zero bias conductance $G_{0}$ of a contact, and the evolution of the excess current $I_{ex}$ measured at large voltage ($eV >> \Delta$), normalised as $I_{ex}R_{n}/\Delta$ where $R_{n}$ is the normal state resistance, both as a function of temperature \cite{Lofwander}. 

\section{Experimental}
\begin{figure}
\includegraphics[width=8cm]{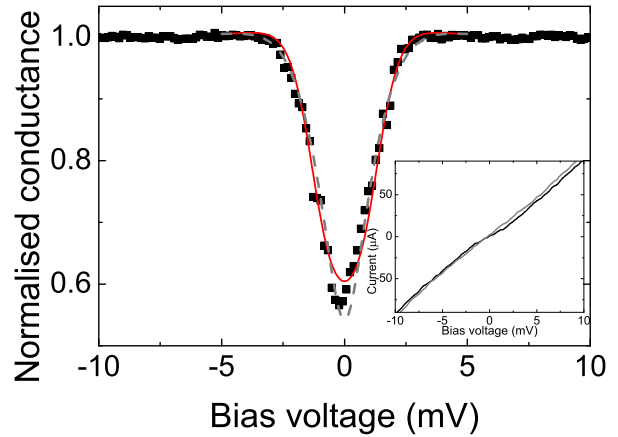}
\caption{Spectrum F at 4.2K normalised by an order 4 polynomial fit (see text) with the BTK fit (red solid line) and the SMM fit (grey dashed line).  Inset shows the IV characteristic at 4.2K (black line) and 7.3 K (grey line).}
\label{fig.1}
\end{figure}

Films of 100nm thickness were grown onto TiO$_{2}$ substrates by chemical vapour deposition (CVD) as described in \cite{AnwarPRB, AnwarAPL}.  Films grown onto Al$_{2}$O$_{3}$ substrates were of varying thickness and are described in ref \cite{Yates, Sousa}. Point contact measurements were taken using mechanically sharpened Pb tips (T$_{c}$ = 7.2K) and using a differential screw mechanism to slowly bring the tip into contact with the sample \cite{Bugoslavsky} in a dewar of liquid helium.  A contact was established at low temperature, and spectra were taken at increasing temperatures until a temperature just greater than the T$_{c}$ of Pb was reached.  The background conductance was found to be temperature dependent meaning that it was not possible to normalise the low temperature spectra using the spectrum taken at T $>$ T$_{c}$.  In order to normalise the spectra, it was found necessary to use a fourth order polynomial curve to fit the experimental curves above the region $|V| \geq 5 $mV. This was performed at each temperature and the data was then divided by this polynomial curve. An example of the resulting normalised curve used in the fitting routine is shown in figure ~\ref{fig.1}.  Spectra were fitted with the SMM and the Mazin modified BTK models. In each case the value for the gap energy, $\Delta$, was fixed to be that of Pb (ie. $\Delta _{0}$ = 1.35meV).  For the SMM case, the fitting parameters were the spin mixing angle, $\theta$, a measure of the interface scattering, $Z_{smm}$ and $r_{s}$.  Following the method of ref \cite{Lofwander}, the value of $P$ was assumed to be fixed to 100\% while the misalignment angle, $\alpha$, was fixed at $\pi /2$ as described in references \cite{Lofwander, Grein} and used in \cite{Lofwander}.  For fitting to the Mazin modified BTK model, the fit parameters were the polarisation, $P$, $Z_{BTK}$ and $r_{s}$. 

\section{Results and Discussion}
\begin{figure}
\includegraphics[width=8cm]{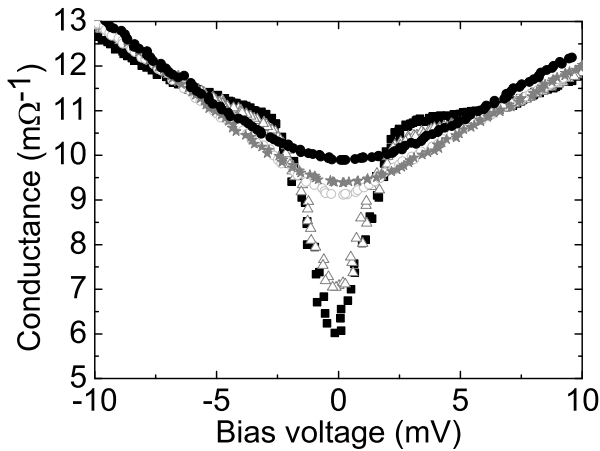}
\caption{Point contact Andreev reflection spectra onto CrO$_{2}$ with a Pb tip at (from bottom) 4.2K, 5.1K, 6.0K, 6.4K, 7.3K.}
\label{fig.2}
\end{figure}

Figure ~\ref{fig.2} shows the temperature dependence of a contact made onto the CrO$_{2}$ thin film.   All spectra showed a suppression of the V $<$ $\Delta$ conductance at T $<$ T$_{c}$ consistent with contacts onto highly spin polarised films.  As the temperature increased, the suppression of the zero bias conductance reduced until the background spectra were obtained at T $\simeq$ 6.5K, indicating the superconducting critical temperature of Pb was suppressed at these interfaces.  In common with previous studies, we will denote the local critical temperature of the contact as T$_{c}^{A}$.  Note that a suppressed T$_{c}$, can result from at least one of two scenarios, either the tips used have been somewhat oxidized or possibly, that there is reasonably strong proximity effect. The latter would have to be taken into account in the fitting model used, as described for example in Strijkers et al.,\cite{Strijkers}, but in our case no evidence for bulk and suppressed gap features (i.e. features associated with two superconducting energy gaps) are observed in the spectra and hence the analysis by Strijkers et al., cannnot be applied \cite{Strijkers}.

An important comparison between the SMM and Mazin modified BTK models is the behaviour of the zero bias conductance as a function of temperature \cite{Lofwander}.  Figure 3 shows the zero bias conductance (normalised to the zero bias conductance at T $\geq$ T$_{c}^{A}$) as a function of temperature. The attraction of this comparison is that the zero bias conductance value is taken straight from the raw data and no fitting is involved. Data on two sets of films are presented; those grown on TiO$_{2}$ substrates and data from films grown on Al$_{2}$O$_{3}$ substrates used in a previous study \cite{Yates}.  The films grown on TiO$_{2}$ have been shown to support LRSTPE \cite{AnwarSUST}, and those grown on Al$_{2}$O$_{3}$ were found to exhibit very high transport spin polarisation values of $\simeq$ 90 \% \cite{Yates, Branford}. Both sets of data show a close to linear decrease of $G_{0}/G_{N}$ down to T/T$_{c}^{A} \simeq$ 0.6.   It is interesting to note that a near linear dependence is expected in the SMM if P = 100\% and $\theta \simeq \pi /2$ whereas the behaviour should show quasi-exponential behaviour in the modified BTK or in the SMM case with $\theta = 0$ (for zero non-thermal broadening and/or zero series resistance), the latter case means that the interface is no longer spin active and the BTK results are recovered as shown by the theoretically generated data also shown in the figure. In our case, the temperature window of our experiments is restricted, nevertheless the trends in the experimental curves suggest that the SMM with $\theta \simeq \pi/2$ fits the majority of data points for films on both types of substrate.   Note that the data can be equally well explained within the BTK model if a polarisation significantly less than 100\% and either a large non-thermal broadening (quite usual for point contact type experiments \cite{Bugoslavsky}) or a large series resistance $r_{s}$  is considered  for all contacts. 

\begin{figure}
\includegraphics[width=8cm]{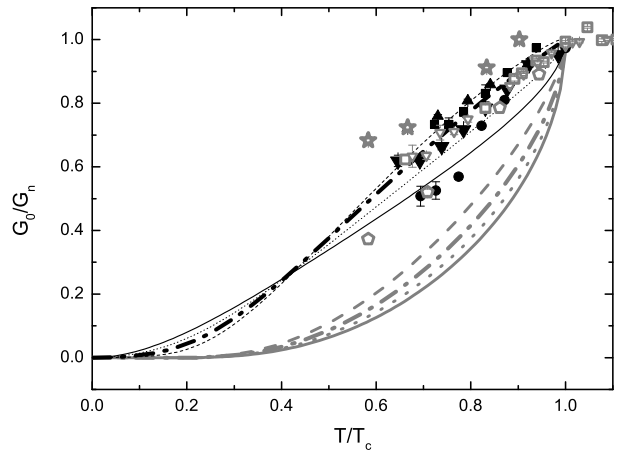}
\caption{$G_{0}/G_{N}$ for contacts grown onto TiO$_{2}$ substrates (closed symbols) onto Al$_{2}$O$_{3}$ substrates (open symbols).  Also shown is the predicted behaviour for $\theta = \pi /2$ (black) and $\theta$ = 0 (grey) for $Z$ = 0.1 (solid line), 0.26 (dotted line), 0.5 (thick dashed line), 1.0 (dashed line).}
\label{fig.3}
\end{figure}

In order to compare the parameters generated by fitting to each model, the lowest temperature spectra of data sets B (4.3K) and F (4.2K) (both on TiO$_{2}$) were each normalised by dividing by the polynomial background.  The resulting fits to both models are shown in figure 1.   On obtaining the fit parameters $P$, $Z_{BTK}$, $r_{s}$ (modified BTK) or $\theta$, $Z_{smm}$, $r_{s}$ (SMM), the $G_{0}/G_{N}$ curve was generated for each data set and model, figure 4(a-d).   Note that for consistency, the data for sets B and F have been replotted in figure 4 as $G_{0}/G_{N\prime}$ where $G_{N\prime}$ is the ‘conductance’ at the zero bias point of the polynomial fit.  The difference between these two normalisation methods is minimal as can be seen by comparing the data for set B to the high temperature normalised data in figure 4c.  Following full fitting of the lowest temperature spectrum and generation of the $G_{0}/G_{N\prime}$ from these parameters, it can be seen that within our temperature window we are unable to differentiate between the models. However, data taken to a lower temperature or the employment of superconductor with a higher critical temperature would facilitate the comparison between models. As it stands, the SMM model assumes the films have 100\% polarisation, which is not unreasonable given the fact they strongly support LRSTPE \cite{AnwarSUST}, but this is not definitive proof. Spin polarised photoemission although restricted to the top few nanometers of the surface would be a useful additional characterisation tool.  Fitting the data within the modified BTK model produces lower values of polarisation (coupled with a high value of $r_{s}$ for contact F.)

\begin{figure}
\includegraphics[width=8cm]{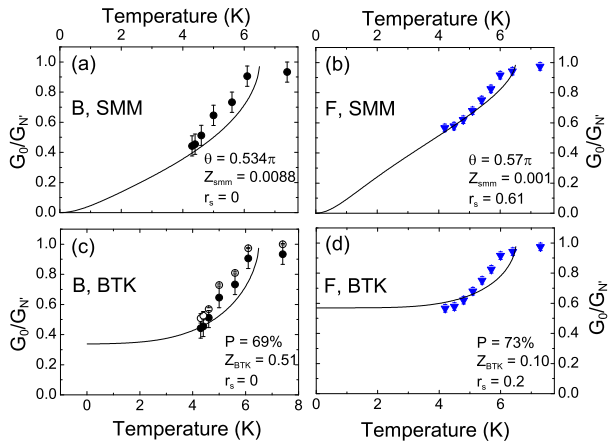}
\caption{$G_{0}/G_{N\prime}$ and the predicted behaviour based on the low temperature fits for the spin mixing model for contact (a) B and (b) F and the modified BTK model for contact (c) B and (d) F, fitting parameters are included in the figure.  The open symbol data in (c) is the data for contact B normalised to T$>$T$_{c}$.}
\label{fig.4}
\end{figure}

Measurement of the excess current was suggested in ref \cite{Lofwander}, as a tool for checking the validity of the fits either to the modified BTK or the SMM. In its simplest form, the deviation of the I-V characteristic from Ohmic behaviour gives the current deficit of the contact, $I_{ex}$. The magnitude of the deficit can then be plotted as $I_{ex}R_{n}/\Delta$ where $R_{n}$ is the normal state resistance.  The value of $I_{ex}R_{n}/\Delta$  is predicted to vary as a function of $Z$ and $P$ in the modified BTK model and $Z$, $P$ and $\theta$ in the spin mixing model \cite{Lofwander}.  The validity of the fit can be established by comparing the value of $I_{ex}R_{n}/\Delta$ at $Z_{fit}$ with the values for the other parameters extracted.  For the contacts shown in figure 4(a-d), the excess current values ($I_{ex}R_{n}/\Delta$) predicted are, for contact B, -0.46 (SMM) and -0.39 (BTK) while for contact F, $I_{ex}R_{n}/\Delta \simeq$ -0.43 (SMM), -0.42 (BTK).  An unfortunate complication in our data is that the I-V curves taken at T$>$T$_{c}$ are also non-Ohmic (see for example the conductance curve at T$>$T$_{c}$ in figure 2  and the IV curve in the inset to figure 1) and it is this non-Ohmic response that dominates the excess current evaluation.  In order to account for this, two methods were applied to approximate $I_{ex}$, firstly the IV at T $<$ T$_{c}$ was subtracted from that just above T$_{c}$. The value of the $I_{ex}$ was then averaged over $|5-10|$mV and the standard deviation from this average was taken as the error.  Secondly, an Ohmic response was taken away from the data at low temperature and compared with the same Ohmic response for data taken at T$>$T$_{c}$, with the final value for  $I_{ex}R_{n}/\Delta$ again being taken as the average over $|5-10|$mV and the standard deviation as the error.  Within the considerable error, the data set matches both predictions from BTK and SMM models.  Therefore although in general this method may provide additional validation, for the particular case of CrO$_{2}$, where there is a strong temperature dependent background,  it does not help distinguish between the models. 

The main result is that there is no detectable difference in behaviour between films grown onto Al$_{2}$O$_{3}$ and those grown onto TiO$_{2}$ using a model independent method.  It has been observed previously \cite{AnwarAPL} that films grown onto TiO$_{2}$ supported the LRSTPE when a magnetically inhomogeneous layer was inserted. It was suggested that the higher degree of magnetic homogeneity achieved for films grown on these substrates may have meant that some components needed to generate the LRSTPE were lacking \cite{AnwarAPL}.  Stimulated by the interpretation within the SMM, we propose that spin mixing is the common ingredient in generating the LRSTPE, and that the spin flip process is furnished by the magnetic disorder (films on Al$_{2}$O$_{3}$) or an extra magnetic layer (Ni for the case of films on TiO$_{2}$) \cite{AnwarAPL}.  It is interesting to speculate that also spin-orbit scattering might be used for this purpose.

In summary we have revisited the spectroscopic information obtained on S/F contacts between Pb and CrO$_{2}$. Although within the limits of our experimental temperature window we are unable to differentiate between models, we have set out the types of experiments that would need to be carried out in order to do so.  No difference is found in the conductance spectra taken on CrO$_{2}$ films grown on different substrates despite their different behaviours in terms of the generation of the long range spin triplet proximity effect.


\acknowledgments
ME acknowledges support from EPSRC under grant reference EP/J010618/1, LFC and KY acknowledge support from EPSRC under grant reference EP/H040048/1.



\bibliography{cro2bib}

\end{document}